\documentclass[aps,prb,twocolumn,showpacs,superscriptaddress]{revtex4}
\usepackage{graphicx}
\usepackage{verbatim}
\usepackage{sidecap}
\usepackage{epstopdf}

\begin{document}

\title{Deviating band symmetries and many-body interactions in a model hole doped iron pnictide superconductor}

\author{L.A. Wray}
\affiliation{Advanced Light Source, Lawrence Berkeley National Laboratory, Berkeley, CA 94720, USA}
\affiliation{Department of Physics, Princeton University, Princeton, NJ 08544, USA}
\author{R. Thomale}
\affiliation{Department of Physics, Princeton University, Princeton, NJ 08544, USA}
\affiliation{Department of Physics, Stanford University, Stanford, California 94305, USA}
\affiliation{Institute for Theoretical Physics, University of W\"{u}rzburg, Am Hubland, D 97074 W\"{u}rzburg}
\author{C. Platt}
\affiliation{Institute for Theoretical Physics, University of W\"{u}rzburg, Am Hubland, D 97074 W\"{u}rzburg}
\author{D. Hsieh}
\affiliation{Department of Physics, Massachusetts Institute of Technology, Cambridge, MA 02139, USA}
\author{D. Qian}
\affiliation{Department of Physics, Shanghai Jiao Tong University, Shanghai 200030, People's Republic of China}
\author{G. F. Chen}
\author{J. L. Luo}
\author{N. L. Wang}
\affiliation{Beijing National Laboratory for Condensed Matter Physics, Institute of Physics, Chinese Academy of Sciences, Beijing 100080, People's Republic of China}
\author{M.Z. Hasan}
\affiliation{Department of Physics, Princeton University, Princeton, NJ 08544, USA}

\begin{abstract}

We present a polarization resolved study of the low energy band structure in the optimally doped iron pnictide superconductor Ba$_{0.6}$K$_{0.4}$Fe$_2$As$_2$ (T$_c$=37K) using angle resolved photoemission spectroscopy. Polarization-contrasted measurements are used to identify and trace all three low energy hole-like bands predicted by local density approximation (LDA) calculations. The photoemitted electrons reveal an inconsistency with LDA-predicted symmetries along the $\Gamma$-X high symmetry momentum axis, due to unexpectedly strong rotational anisotropy in electron kinetics. We evaluate many-body effects such as Mott-Hubbard interactions that are likely to underlie the anomaly, and discuss how the observed deviations from LDA band structure affect the energetics of iron pnictide Cooper pairing in the hole doped regime.

\end{abstract}


\date{\today}

\maketitle

\section{Introduction} %I

Despite intensive efforts, the electronic structures underlying pnictide high-T$_c$ superconductivity continue to include a number of areas of experimental uncertainty. This is largely due to the fact that iron pnictides are multi-band correlated systems with a complicated interplay of spin, orbital, and lattice degrees of freedom \cite{discovery,spinOrder,YildirimPhonon,KreyssigCa,DingOrbitals,chargeAndLattice,MazinSpinMech,MYiDetwinning,MazinReview}. Here we explore the low energy electronic structure of the optimally doped superconductor Ba$_{0.6}$K$_{0.4}$Fe$_2$As$_2$ (T$_c$=37K), using angle-resolved photoemission spectroscopy (ARPES) measurements performed with high-symmetry experimental geometries to provide a rigorous basis for understanding reflection symmetries and electron kinetics of the three hole-like bands found in the Brillouin zone (BZ) center. Strong similarities are known to exist between renormalized paramagnetic local density approximation (LDA) predictions and the experimentally observed electronic structure of these bands \cite{DingOrbitals,WrayBaK,MingYi,Borisenko,DingBS,FengGap,DingKink,DingGap,Ding3DGap,Mannella3D,FeFephase_Brouet}, however many-body interactions and local symmetry breaking create the potential for significant deviations \cite{MYiDetwinning,Borisenko}. In this study, careful control of the measurement geometry is used to identify a region of band structure with clear discrepancy between photoemission data and electron symmetries predicted in LDA.

\begin{figure}
\includegraphics[width = 8.7cm]{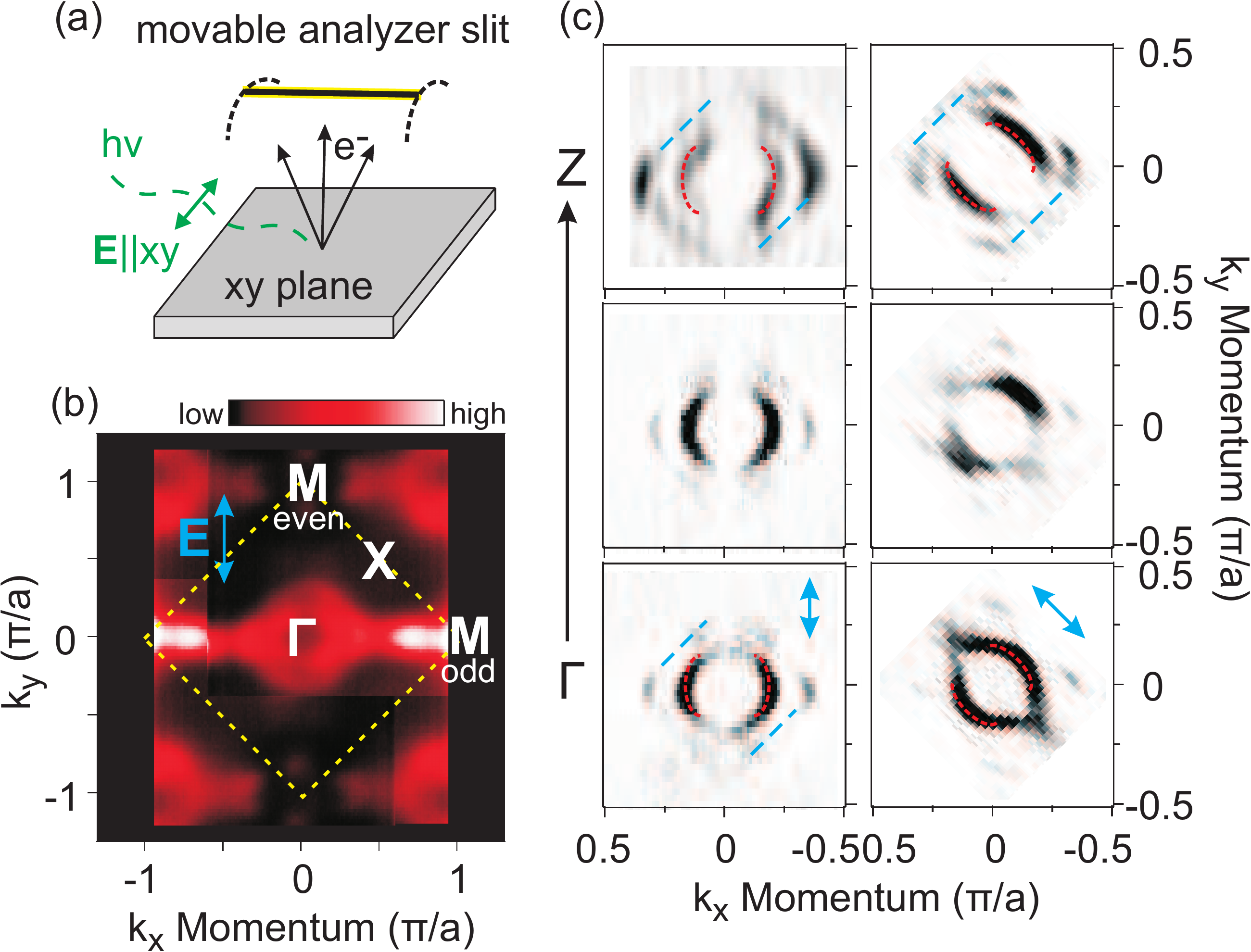}
\caption{\label{fig:measIntro}{\bf{Fixed polarization Fermi surface mapping}}: (a) Polarization is kept fixed by moving the analyzer rather than the sample during Fermi surface mapping, and measurements along arbitrary momentum directions are derived from the resulting constant-polarization data matrix. (b) High symmetry points in the 2D Brillouin zone are labeled on a Fermi surface map. (c) The three dimensional Fermi surface is constructed from second derivative Fermi surface images measured at momenta spanning the k$_z$-axis Brillouin zone. Polarization is parallel to the y-axis for the left column and to the [x-y]-axis for the right column. A red guide to the eye traces intensity associated with the $\alpha_2$ band defined in Fig. 2, and portions of the outermost FS contour are indicated in blue. Linear dimensions for these guides to the eye are increased by 10$\%$ in the Z-point plane (k$_z$=$\pi$) relative to the BZ center.}
\end{figure}

Our measurements show that electrons in the inner pair of hole bands have interchanged reflection symmetries relative to LDA predictions along the $\Gamma$-X momentum axis, caused by an unexpectedly large rotational anisotropy in electron kinetics. The outermost hole band also shows significant inconsistencies with LDA band kinetics realizing a much larger density of states at the Fermi level than is attributed in most first principles based predictions. We conclude the paper by reviewing many-body effects that may account for the discrepancies with LDA, and discussing the implications of our measurements with respect to the energetics of Cooper pairing.

\begin{figure}
\includegraphics[width = 8cm]{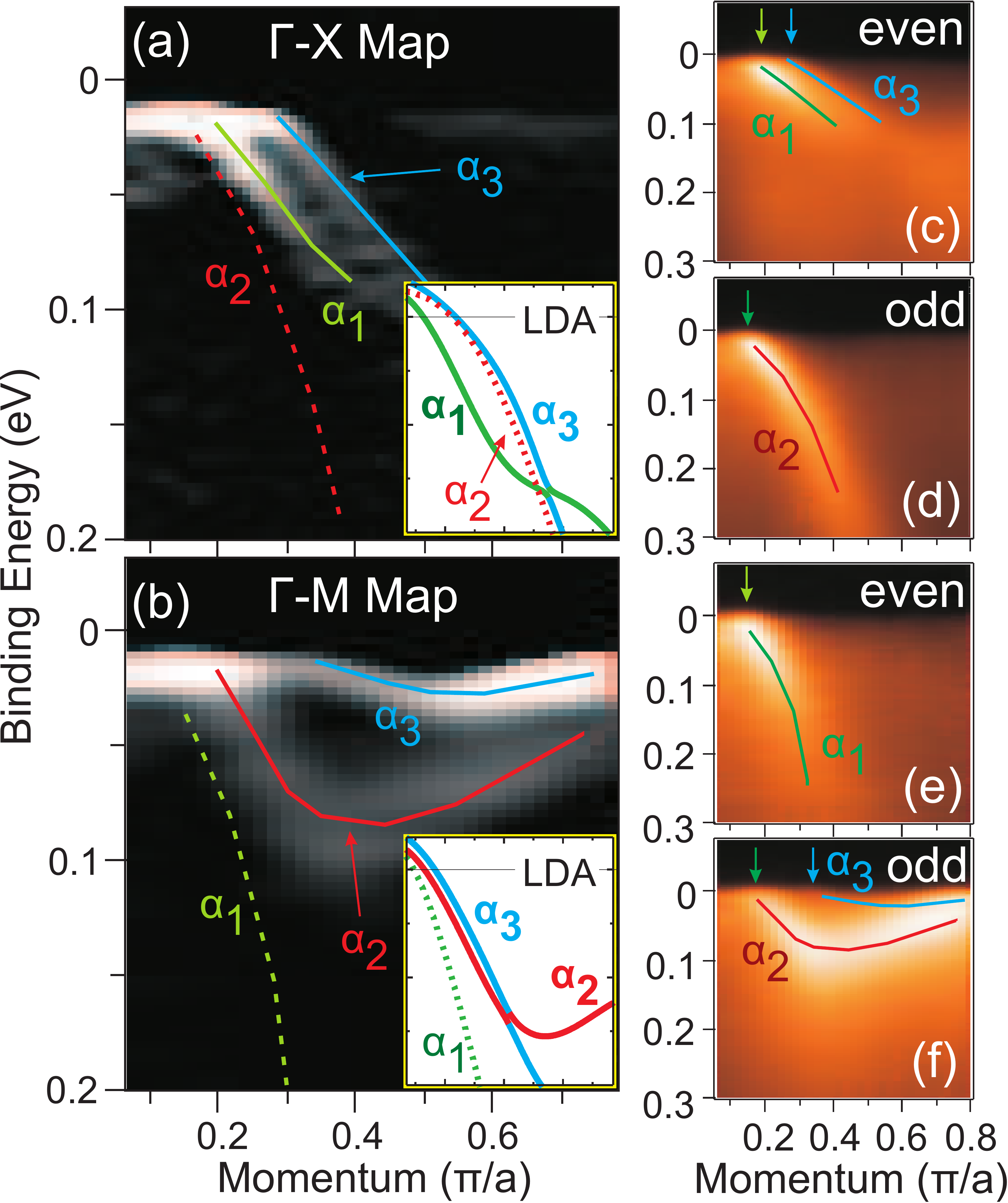}
\caption{\label{fig:straightCuts}{\bf{High-symmetry mapping of hole-like bands}}: (a-b) Second derivative images (SDI) of ARPES measurements along the $\Gamma$-X and $\Gamma$-M axes are traced with the approximate dispersion of all three bands found in our measurements. Panel (a) is measured with $even$ polarization and panel (b) with $odd$ polarization. Dispersions over the same energy/momentum window from LDA are shown in the inset using a 50$\%$ renormalization factor as in Ref. \cite{DingBS}, with dashed lines used to represent bands that should not be visible because they are in the incorrect symmetry sector for photoemission. (c-d) ARPES images along the $\Gamma$-X axis are shown for in-plane polarization oriented 0$^o$ ($even$) and 90$^o$ ($odd$) relative to the measurement axis. (e-f) ARPES images are shown for the $\Gamma-M$ axis.}
\end{figure}

Hole doped superconducting Ba$_{1-x}$K$_x$Fe$_2$As$_2$ has been widely studied due to the extremely high sample quality available, as noted in angle resolved photoemission spectroscopy (ARPES), scanning tunneling microscopy ($\sim$1$\AA$ rms surface roughness) and magnetic susceptibility measurements \cite{WrayBaK,DingBS,Borisenko,DingKink,Ding3DGap}. The location of As atoms at the optimally doped composition of Ba$_{0.6}$K$_{0.4}$Fe$_2$As$_2$ is 35.3$\pm$0.2 degrees out of plane relative to their nearest neighbor Fe atoms \cite{KimberAsAngle}, extremely close to the value of $\sim$35$^o$ thought to be ideal for superconductivity \cite{KurokiPnictogenHeight}. These factors correlate with the large superconducting critical temperature (T$_c$$\sim$37K) and suggest that the system is a model realization of hole-doped iron pnictide superconductivity.

\begin{figure*}[]
\includegraphics[width = 16cm]{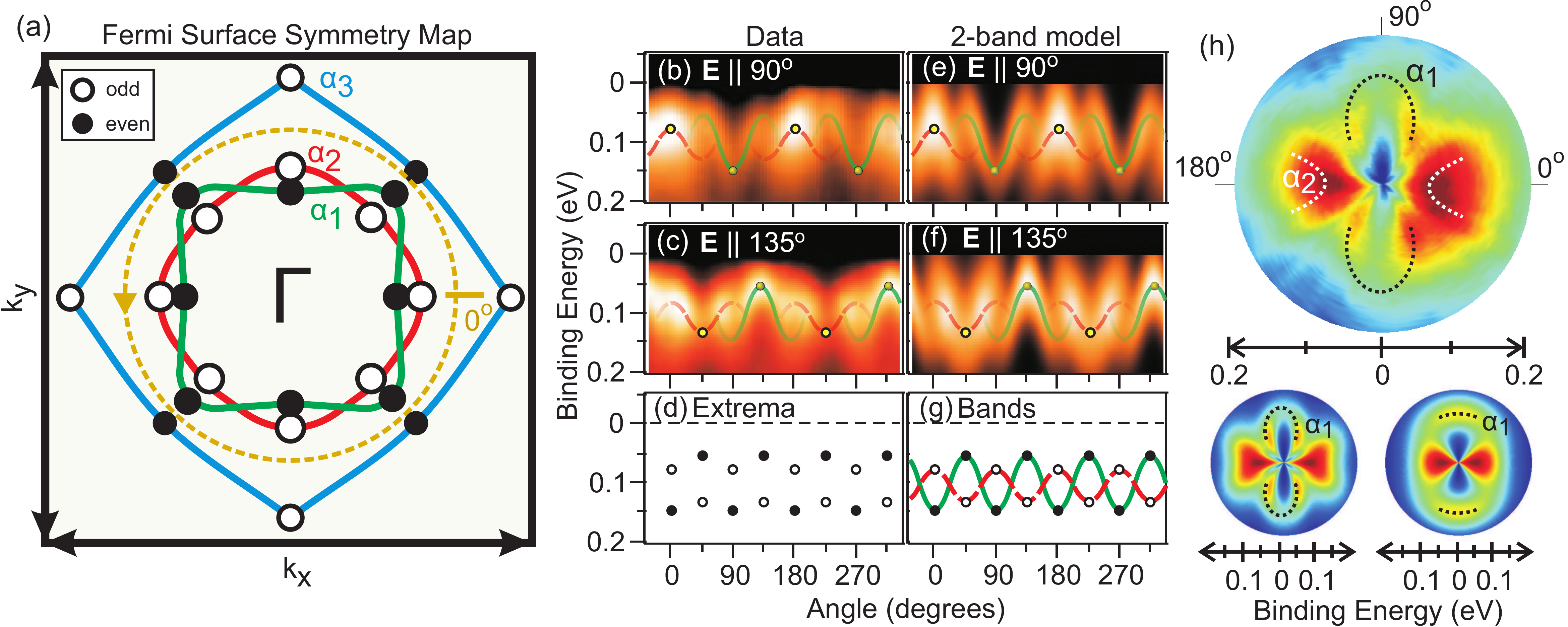}
\caption{\label{fig:circleCuts}{\bf{Circular measurements show intertwined bands}}: (a) A Fermi surface diagram shows how the band symmetries identified in Fig. 2 can be reconciled with the LDA band structure. The measured (dominant) reflection symmetry for each band at momenta along the high symmetry axes is labeled with respect to radial measurements intersecting the Brillouin zone center. ARPES intensity along the circular orange contour traced in panel (a) is measured for polarization along (b) the y-axis and (c) the $\hat{y}-\hat{x}$ -direction. (d) The energies of the (black circles) $\alpha_1$ and (white circles) $\alpha_2$ bands are obtained based on the point of maximum intensity under symmetry conditions for which only one band is distinctly visible (0$^o$ and 90$^o$ from the polarization axis). (e-g) ARPES intensity is predicted within a 2-band k.p model of the 3d$_{xz}$ and 3d$_{yz}$ orbitals, fitted from the band energy extrema identified in panels (b-d). (h) The data from panel (b) are plotted as a polar image, for comparison with (bottom left) the intersecting 2-band model from panel (e) and (bottom right) a scenario in which the $\alpha_1$ and $\alpha_2$ bands are rotationally isotropic. Intensity from the $\alpha_1$ band has been traced in black as a guide to the eye.}
\end{figure*}

\section{Methods and Definitions} %II

Single crystals of Ba$_{0.6}$K$_{0.4}$Fe$_2$As$_2$ (T$_c$=37 K) were grown using the self-flux method \cite{ChenGrowth}. ARPES measurements were performed at the Advanced Light Source beamline 10.0.1 using 34-51 eV photons with better than 15 meV energy resolution and overall angular resolution better than 1$\%$ of the Brillouin zone (BZ). Samples were cleaved and measured at temperatures below 15 K, in a vacuum maintained below 8$\times$10$^{-11}$ Torr.

High symmetry points in the Brillouin zone are labeled on a Fermi surface image in Fig. 1(b), with the x- and y-axes oriented parallel to the nearest neighbor iron-iron direction ($\Gamma$-M momentum axis). The lattice has mirror reflection symmetry along certain high symmetry axes, causing all electronic states at momenta along those axes to have either an $even$ ($\tilde{R}|\Psi\rangle$=$+1$$|\Psi\rangle$) or $odd$ ($\tilde{R}|\Psi\rangle$=$-1$$|\Psi\rangle$) eigenvalue of an appropriate reflection symmetry operator $\tilde{R}$ unless a symmetry of the electronic system has been broken. Mirror eigenvalues discussed throughout this paper are measured with respect to mirror planes that contain the z-axis and the reciprocal vector of the observed electron momentum (i.e. the x-axis, when measuring along the $\Gamma$-M momentum axis). For measurements along the $\Gamma$-X axis, the mirror plane goes through Fe and As atoms and lies along the main diagonal of the Fe lattice. Along the $\Gamma$-M axis, the mirror plane goes through nearest neighbor As atoms, and does not intersect Fe atoms \cite{FeFephase_Brouet}. Application of the reflection operator $\tilde{R}$ reflects the spatial wavefunction index normal to this mirror plane, as discussed in earlier papers \cite{DamascelliReview,FeFephase_Brouet}. Within ARPES data, only states with $even$ mirror symmetry are observed when incident photon polarization is within the mirror plane, and only states with $odd$ symmetry are observed when polarization is oriented along the axis normal to the mirror plane (e.g. see Eq. [22] of Ref. \cite{DamascelliReview}). For the sake of simplicity, incident photon polarization is oriented purely along the high-symmetry axes of the x/y pnictide plane for all measurements (e.g. \textbf{E}$\parallel$$\hat{y}$ or \textbf{E}$\parallel$$\hat{y}-\hat{x})$. Polarization is kept fixed by adopting the unconventional approach of moving the analyzer rather than the sample to map outgoing momenta covering the full 2D Brillouin zone (diagram in Fig. 1(a)).

For a highly two dimensional multi-band system with strongly anisotropic photoemission matrix elements, it is challenging to fully rule out surface and domain effects, which can add bands, reshape self energy contours and potentially mask bulk electron symmetries. However, several factors suggest that the measurements presented here are representative of bulk electronic properties. The linear dimensions of second derivative (SDI) Fermi surface contours appear to increase by roughly 10$\%$ as momentum is varied across the Brillouin zone along the z-axis (Fig 1(c)), a feature not expected for surface states or resonances. Out of plane momentum is varied by tuning the incident photon energy, and estimated using an inner potential of 15eV consistent with previous studies \cite{Mannella3D,FengGap,Ding3DGap}. We caution however that these measurements in Fig. 1(c) are not ideal for quantitative study of the outermost hole pocket. Due to convolution with the Fermi function, Fermi level second derivative images significantly deemphasize (and potentially shift) the energetically sharp outer hole pocket relative to similar SDI measurements below the Fermi level shown in other figures. More quantitative evidence against the existence of highly domain-dependent band structure will be found in the analysis of Fig. 2-3, where a one-to-one correspondence is established between LDA reflection symmetries and the experimental measurements. Earlier experiments have found circular polarization matrix elements for this compound to be consistent with bulk emission \cite{DingOrbitals,surfacAndCircPol}.

\section{An experimental map of band structure and mirror symmetries} %III

Measuring the Fermi surface with a single fixed incident photon polarization (Fig. 1(b-c)) shows only two bands of intensity surrounding the BZ center, and does not clearly resolve a third circular hole pocket expected from calculations. However, the existence of a third band can be seen from strong ninety degree rotational anisotropy in the innermost Fermi contour when polarization is directed along the $\Gamma$-X axis ([x-y] axis), with a larger Fermi momentum found along the direction parallel to the incident photon polarization (Fig. 1(c,right)). In this section, we will show that this elongation is due to a band symmetry inversion not expected from LDA, but which can be explained from the lowest order effect of d-orbital strong correlation interactions. Measurements presented in Fig. 2-4 of this paper are constrained to k$_z$$\sim$4 rlu, as this is the plane in which LDA suggests that it is \emph{most difficult} for the symmetry inversion to occur.

To observe the bands more clearly, we perform measurements along both $\Gamma$-X and $\Gamma$-M high symmetry directions with photon polarization parallel and perpendicular to the mirror planes defined in Section II for these axes (Fig. 2(c-d,e-f)). These geometries each selectively suppress at least one band by isolating the $even$ or $odd$ symmetry bands, making it easier to separately resolve the remaining bands and map their dispersions close to the Fermi level. Second derivative images in Fig 2(a,b) are used to enhance contrast under geometries for which two bands are visible simultaneously, and are overlaid with the dispersion of the missing band. All bands along the $\Gamma$-X direction follow typical hole-like dispersions, however the two outermost bands in the $\Gamma$-M cut fold upwards as they approach the M-point. The outermost band has a much weaker photoemission signal than the inner two, but is also less broad in energy causing it to appear with greater intensity in second derivative images. The inner two bands have peak widths at half maximum that range from $\delta$E$\sim$90-200meV depending on momentum and binding energy. Based on the factor of $\gtrsim$10 increase in intensity of the outermost band relative to the inner features in second derivative images, we can very roughly estimate that it has an energy width $\delta$E$\sim$30$\pm$15 meV at shallow binding energies (30meV$\sim$100meV/$\sqrt{10}$).

The reflection symmetries and Fermi momenta seen in Fig. 2 are summarized on a Fermi surface schematic in Fig. 3(a), and the individual bands are labeled as $\alpha_1$, $\alpha_2$ and $\alpha_3$ in reference to the LDA band structure. Along the $\Gamma$-X direction (diagonal axis), the reflection symmetry probed by ARPES is evaluated by considering a mirror plane that intersects Fe atoms, meaning that the reflection symmetry of individual iron d-orbitals with respect to that mirror plane will match the symmetry character of bands with which they hybridize. The three bands predicted by LDA have $even$ ($\alpha_1$, strong 3d$_{xz}$+3d$_{yz}$ orbital occupation), $odd$ ($\alpha_2$, strong 3d$_{xz}$-3d$_{yz}$) and $even$ ($\alpha_3$, strong 3d$_{xy}$) symmetry in this direction. Our data also show two $even$ bands and one $odd$ symmetry band, meaning that the innermost, $odd$ symmetry band can be identified with $\alpha_2$ from LDA. This assignment requires that the $\alpha_1$ and $\alpha_2$ bands be inverted relative to their appearance in LDA simulations, which place $\alpha_1$ at larger binding energy than $\alpha_2$ near the Fermi level.

Along the $\Gamma$-M momentum axis (k$_x$), reflection symmetry is determined from a mirror plane that intersects As atoms but not Fe, meaning that without performing a calculation, it is not straightforward to see which Fe d-orbitals will be allowed to hybridize with one another, and what reflection symmetry sector they will be associated with. According to LDA calculations, the $\alpha_1$ band has $even$ symmetry with predominant 3d$_{xz}$ character, $\alpha_2$ has $odd$ symmetry with strong 3d$_{yz}$ occupation, and $\alpha_3$ has $odd$ symmetry and nearly 100$\%$ 3d$_{xy}$ orbital character at the Fermi level. The band dispersion that most closely matches LDA, with $\alpha_1$ dispersing downwards and $\alpha_2$ and $\alpha_3$ bending upwards (see Fig-2(k)), also matches these dominant d-orbital reflection symmetries in this part of the BZ \cite{FeFephase_Brouet}.

\begin{figure*}
\includegraphics[width = 13cm]{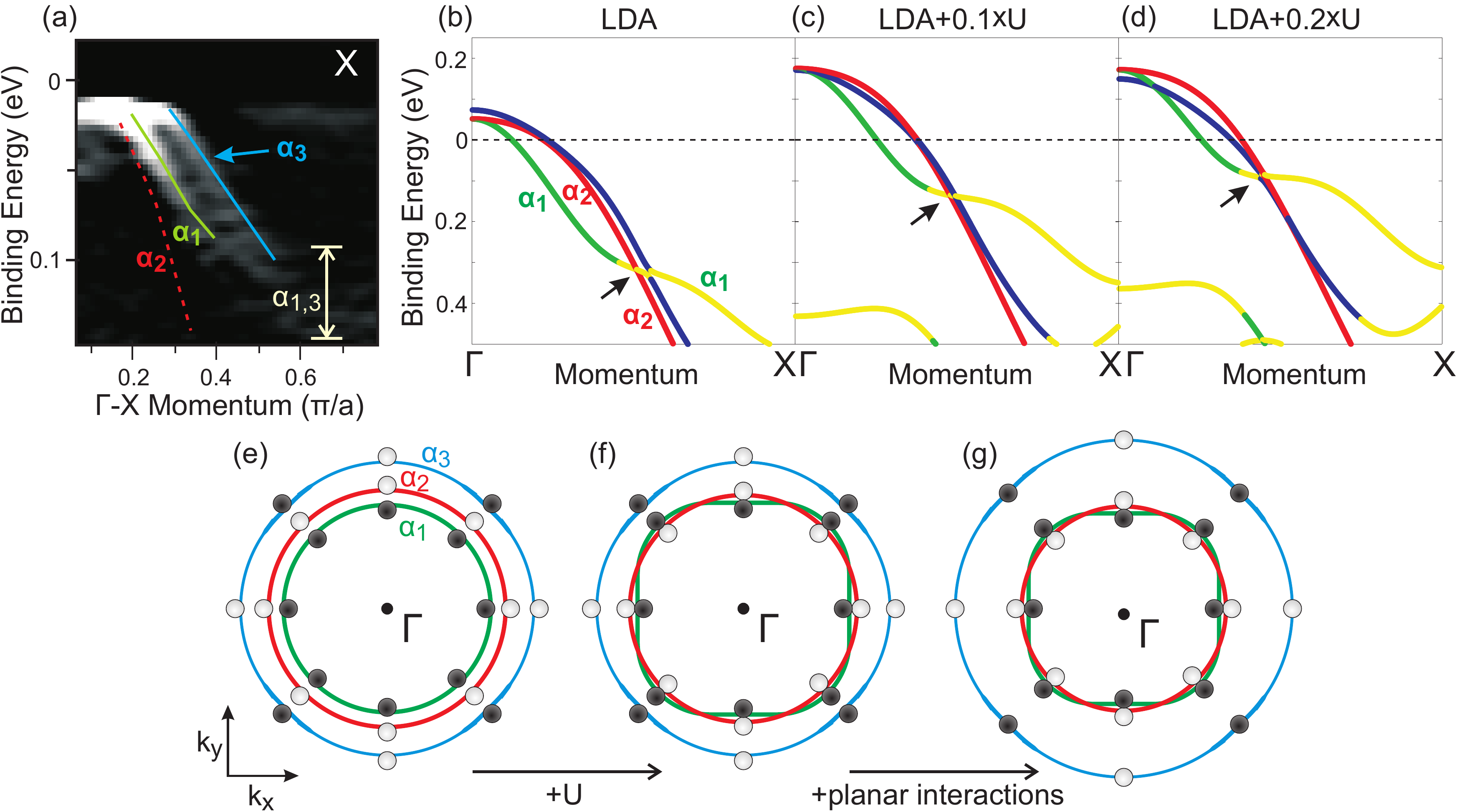}
\caption{\label{fig:summaryFig}{\bf{Reshaping the Fermi surface}}: (a) Dispersion of hole-like bands along the $\Gamma$-X direction is reproduced from Fig. 2(b). (b) LDA band structure is color-coded based on the dominant orbital symmetry, and the $\alpha_1$/$\alpha_2$ symmetry inversion point is indicated with an arrow (yellow-red crossing). The color code is blue for d$_{xy}$ (orbital filling n$_{d_{xy}}$=0.5500), green for the $\alpha_1$ chirality of d$_{xz/yz}$ and red for the $\alpha_2$ chirality (orbital filling n$_{d_{xz/yz}}$=0.5815), and yellow for d$_{3z^2-r^2}$ (orbital filling n$_{d_{3z^2-r^2}}$= 0.7301). (c-d) The LDA band structure is modified with a small Mott-Hubbard perturbation as described in the text, with Fermi level set to maintain constant carrier density. Cartoons in (e-g) illustrate how this and other effects may modify the Fermi surface: (e) Bare LDA results give similar radii for all three Fermi pockets. (f) Perturbing orbital energies with on-site Coulomb interactions increases the anisotropy of $\alpha_1$ and $\alpha_2$, consistent with the ``intertwined" bands observed experimentally. (g) Strong in-plane intersite interactions and correlations may be responsible for the enlarged $\alpha_3$ Fermi surface and density of states. Dark circles represent even reflection symmetry and light circles represent odd reflection symmetry, as defined for Fig. 3(a).}
\end{figure*}

\begin{figure}
\includegraphics[width = 8cm]{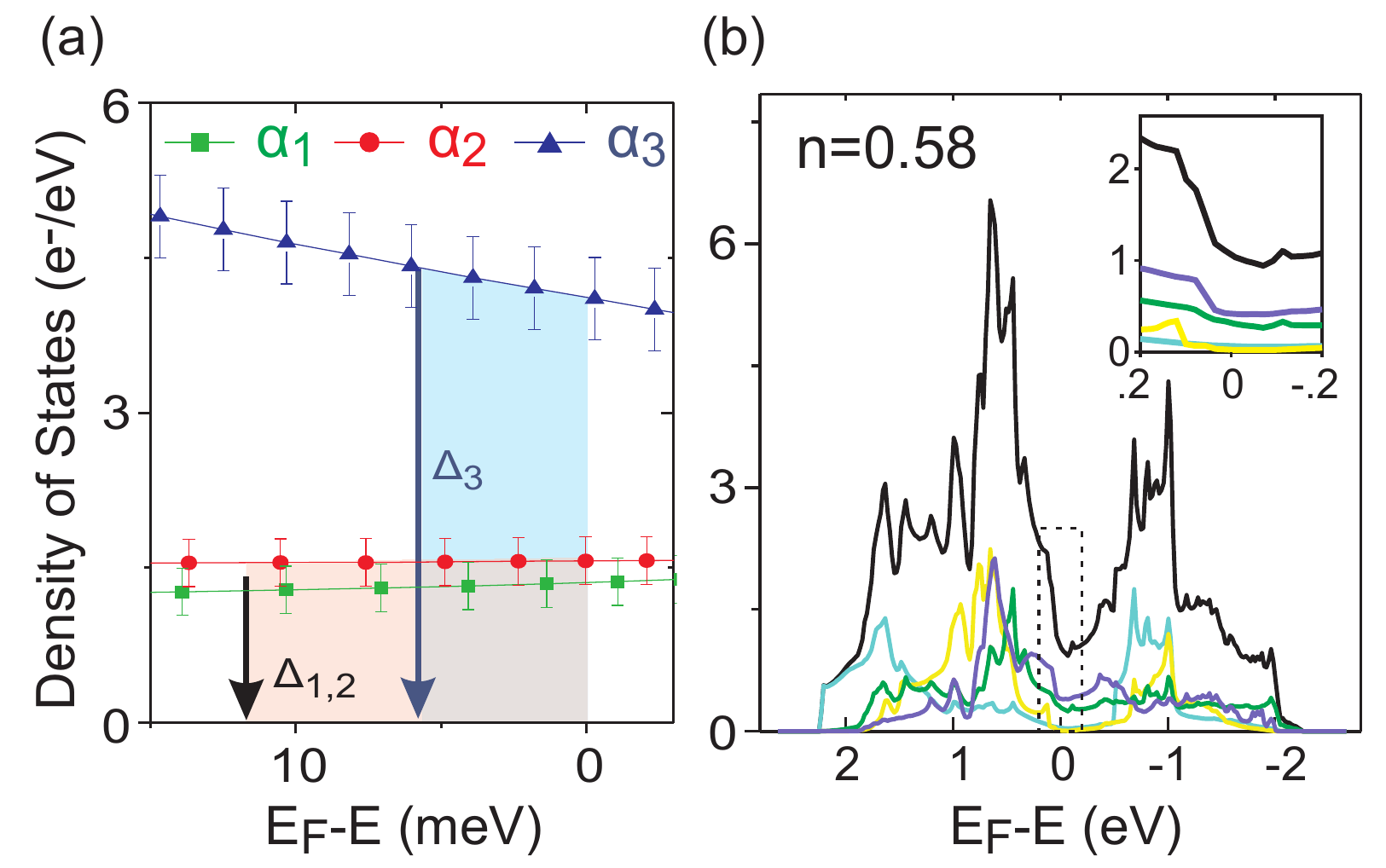}
\caption{\label{fig:DOSandLDA}{\bf{Bands and orbitals of superconducting electrons}}: (a) Density of states of the $\alpha_1$, $\alpha_2$ and $\alpha_3$ bands near the Fermi level is obtained from fitting the experimental data in Fig. 2. The approximate gap energies for these bands are labeled based on earlier studies ($\Delta_1$$\sim$$\Delta_2$$\sim$13meV, $\Delta_3$$\sim$6meV \cite{DingGap,WrayBaK}). (b) The orbital-resolved 3D integrated density of states (DOS) is obtained from LDA, and an inset shows electrons near the Fermi level. Total filling is 0.58, consistent with the degree of potassium doping. As in Fig. 4(b-d), the color code is blue for d$_{xy}$, green for d$_{xz/yz}$, yellow for d$_{3z^2-r^2}$, and cyan for d$_{x^2-y^2}$. The total density of states is shown in black.}
\end{figure}

Reconciling the dispersion and symmetries observed for $\alpha_1$ and $\alpha_2$ along the $\Gamma$-M and $\Gamma$-X axes can be achieved by intertwining the LDA bands as drawn in Fig 3(a). Such an intertwined Fermi surface requires that the bands have much greater rotational anisotropy than is expected from LDA, a factor that has not been evaluated in previous studies that compared LDA band structure with photoemission data \cite{DingBS,MingYi,Quazilbash}. We have therefore performed several additional measurements to confirm that the $\alpha_1$ and $\alpha_2$ bands intersect. By examining ARPES intensity along a circular momentum contour (dashed line in Fig. 3(a)) inside the $\alpha_3$ Fermi surface, we can observe the anisotropy and polarization matrix elements of each band shown in Fig. 3(b-c). We find that one of these bands is always suppressed at a 0$^o$ or 90$^o$ angle to the incident polarization, meaning that the binding energy of each band is easy to identify from the point of maximum intensity in these geometries as plotted in Fig. 3(d). Connecting the dots with intersecting sinusoidal functions leads to the band anisotropies traced as guides to the eye in Fig. 3(b-c).

The $\alpha_1$ and $\alpha_2$ bands have been described in previous literature with a model based on the 3d$_{xz}$ and 3d$_{yz}$ orbitals that generates ARPES matrix elements with respect to the incident polarization angle of sin$^2$($\theta$-$\theta_{pol}$) for $\alpha_1$ and cos$^2$($\theta$-$\theta_{pol}$) for $\alpha_2$ \cite{Ying2Band,DingOrbitals}. Fitting this low order model to the binding energies and symmetries identified from Fig. 3(b-d) \cite{SMo} gives the simulated ARPES images shown in Fig. 3(e-f), which show how intertwined $\alpha_1$ and $\alpha_2$ bands may be expected to appear along a circular contour, and provide a reasonable match with the experimental data. Within this model, hybridization is not allowed between $\alpha_1$ and $\alpha_2$, consistent with the lack of a visible hybridization gap where bands intersect $\sim$20 degrees from the $\Gamma$-M axis in Fig. 3(b). The intensity pattern expected from a contrasting scenario in which $\alpha_1$ and $\alpha_2$ have isotropic dispersions similar to first principles predictions is shown in a polar plot (Fig. 3(h), lower right). The key difference between these scenarios in the polar representation is that in the symmetry inverted case revealed by our data, the relatively anisotropic $\alpha_1$ band can be seen oscillating in a `U'-like contour at the top and bottom of the polar chart.

\section{Many-body physics and Fermi-level electron kinetics} %IV

The symmetry analysis and direct measurements of circular anisotropy in Section III both reveal that the $\alpha_1$ and $\alpha_2$ bands are inverted in energy relative to LDA predictions along the $\Gamma$-X axis. To understand how this can come about, we have calculated the perturbative effect of strong correlation interactions, using Mott-Hubbard correlation terms (`U' matrix) obtained from random phase approximation numerics in Ref. \cite{Umatrix}. In LDA calculations, the $\alpha_1$ and $\alpha_2$ bands become inverted by crossing well below the Fermi level, at the point indicated with an arrow in Fig. 4(b). Using an LDA+U formalism to apply a weak Mott-Hubbard perturbation (0.1$\times$U) to the orbital energies demonstrates that Mott-Hubbard interactions have the immediate effect of driving the symmetry inversion point up towards the Fermi level, consistent with our data (Fig. 4(c-d)). The most significant factor in this change is an upward shift in the energy of the 3d$_{3z^2-r^2}$ orbital, which mixes with the $\alpha_1$ band along the $\Gamma$-X axis and has approximately 20$\%$ larger occupation than any other orbital. A large electronic occupation raises the energy of 3d$_{3z^2-r^2}$ relative to other d-orbitals when correlation effects are considered. The 3d$_{3z^2-r^2}$ orbital does not hybridize with $\alpha_1$ along the $\Gamma$-M axis, thus this effect can only invert the $\alpha_1$ and $\alpha_2$ bands close to the $\Gamma$-X axis. An additional corrective factor is needed to obtain the Fermi momentum observed in our data for the $\alpha_3$ band, possibly because the lobes of the 3d$_{xy}$ orbital that it is derived from are oriented within the pnictide plane and point towards nearest-neighbor As atoms, leading to strong in-plane electronic interactions \cite{Umatrix}. This conjecture and other band structure corrections considered in this section are summarized in Fig. 4(e-g).

The fact that the BZ center photoemission matrix elements mapped in this study show no sign of mixing between the even and odd symmetry sectors is also noteworthy with respect to the many-body electronic environment. Spin order and other types of many-body structure that are important to models of pnictide compounds can potentially break reflection symmetry or cause additional bands to fold into the BZ center from other parts of momentum space \cite{spinOrder}, particularly if they manifest with a length scale greater than electronic quasiparticle coherence.

In particular, the stripe-like type-1 antiferromagnetic fluctuations known to exist in BaFe$_2$As$_2$ and other iron pnictides \cite{spinOrder,MazinSpinMech} can directly break reflection symmetry for photoemission, and are thought to couple to the $\alpha_3$ band of Ba$_{0.6}$K$_{0.4}$Fe$_2$As$_2$ \cite{Borisenko}. Two other many-body effects that could introduce additional bands near the BZ center include the $\Gamma$-X periodicity surface reconstruction known from scanning tunneling microscopy measurements \cite{STMreconstruct}, and hypothetically, short-range checkerboard antiferromagnetic order such as is found in 2D cuprates. The loss of z-axis translational symmetry from cleavage does not directly alter the reflection symmetries studied, but could introduce additional surface-localized bands. Thus, the lack of significant reflection symmetry breaking or additional bands in BZ center photoemission along both $\Gamma$-M and $\Gamma$-X axes provides a limiting condition for models of these effects in superconducting pnictides.

\section{Band symmetries and superconducting energetics} %V

The polarization resolved measurements in Fig. 2-3 give a clear view of individual band dispersions and anisotropy near the Fermi level, which are critical components in any model of Cooper pairing. Tracing these dispersions with a low-order Taylor series provides an estimate of the band-resolved density of states (DOS), which we find to be larger and differently contoured than the DOS predicted by LDA (Fig. 5(a-b)). The fraction of the superfluid ground state derived from each band is coarsely approximated by multiplying density of states by the superconducting gap size, giving the volumes shaded in Fig. 5(a). Density of states near the Fermi level from the $\alpha_1$ and $\alpha_2$ bands is approximately 2.9$\pm$0.3 e$^-$/eV and the low temperature superconducting gap function is $\Delta_{SC}$$\sim$13$\pm$2 meV \cite{WrayBaK,FengGap,DingGap,Ding3DGap}, giving a superfluid volume of n$_S$=0.038$\pm$0.010e$^-$ per unit cell from those two bands combined (0.038e$^-$= 2.9e$^-$/eV$\times$0.013eV). This value must be divided by two to obtain the number density of Cooper pairs. Though the $\alpha_3$ band is close to the nodal ring in the s$_{\pm}$ superconducting order parameter and has a relatively small superconducting gap of $\sim$6$\pm$2 meV \cite{WrayBaK,FengGap,DingGap,Ding3DGap}, its high 4.3$\pm$0.4 e$^-$/eV density of states causes it to become a significant component of the superfluid with n$_S$=0.026$\pm$0.011e$^-$ superconducting electrons per unit cell (0.026e$^-$=4.3e$^-$/eV$\times$0.006eV). Even though we are not considering electrons near the M-point, as band structure there is more difficult to measure and subject to some controversy \cite{MingYi,Borisenko,DingBS}, there is broad consensus that a great majority of charge carriers in this highly hole-doped compound are associated with the hole pockets at the $\Gamma$ point, and these are expected to constitute most of the superfluid volume. The energetic contribution to the superconducting ground state is estimated from the product of half the superfluid density and the gap size, yielding E$_{1,2}$=0.25$\pm$0.1 meV per unit cell from $\alpha_1$ and $\alpha_2$ (combined), and and E$_3$=0.08$\pm$0.06 meV per unit cell from $\alpha_3$ electrons. If we very roughly estimate that the minimum coherence area for a Cooper pair is on the order of 10 lattice sites \cite{WrayBaK,Hc2}, the temperature scale obtained from these energies is T$_\theta$$\sim$40K (10$\times$(E$_{1,2}$+E$_3$)/k$_B$$\sim$10$\times$0.35meV/k$_B$=41K), which is appropriate for the superconducting critical temperature of T$_c$=37K within the bounds of qualitative comparison.

The large superfluid density estimated from our data for the weakly gapped $\alpha_3$ band is consistent with thermal conductivity studies that have observed a high density of nearly ungapped hole-like states in the superconducting state \cite{JoeThermalHall}. The extremely large DOS and squared-off Fermi surface contour of the $\alpha_3$ band both present significant deviations from first principles predictions, and are a critical element determining the superconducting order parameter \cite{ronnyFeAsvsFeP}. It is particularly interesting that the dispersion of $\alpha_3$ in our data covers a very small $\sim$20 meV energy range, which is not large relative to any of the widely considered many body interactions present in iron pnictides, including the superconducting coherence energetics and spin interactions. In numerical studies, a similarly squared-off Fermi contour and large density of states are typically only obtained for $\alpha_3$ when considering highly hole doped samples (e.g. KFe$_2$As$_2$), where direct nesting between $\alpha_3$ electrons is expected to generate an instability for checkerboard antiferromagnetism and d-wave superconductivity \cite{ronnydWave}.

%Check language:
Inversion of $\alpha_1$ and $\alpha_2$ occurs within 0.1eV of the Fermi level for approximately $\pi$ radians (half) of momentum space surrounding the BZ center (Fig. 3(b-g)), a region of momentum space that is integral to the energetics of superconductivity due to the large $\alpha_1$ and $\alpha_2$ superconducting order parameter. The $\alpha_1$ and $\alpha_2$ bands derive primarily from the degenerate 3d$_{xz}$/3d$_{xz}$ orbital basis, and their relative energies near the $\Gamma$-point are defined by the lowest order dynamical term in k.p theory models \cite{Ying2Band}. Observation of an inversion at small momentum is thus surprising, and implies that a significant correction to standard paramagnetic LDA may be required to accurately model the band symmetries underlying superconducting energetics near the $\Gamma$-X axis. Our LDA+U simulations in Fig. 4(c-d) suggest that the large anisotropy of the $\alpha_1$ and $\alpha_2$ hole bands is related to hybridization with the 3d$_{3z^2-r^2}$ orbital, which may therefore have a strong presence in Fermi level electron symmetries.

In summary, we present a polarization resolved ARPES study of the Fermi surface and band structure in optimally doped Ba$_{1-x}$K$_x$Fe$_2$As$_2$, revealing key ingredients for future numerical models that will comprehensively address the short range symmetry breaking and moderate charge correlations that characterize the pnictide superconductors at the midpoint between conventional superconductivity and the more theoretically intractable strongly correlated regime of cuprate-type superconductivity. We observe the dispersion of three hole-like Fermi sheets surrounding the $\Gamma$-point and use reflection symmetry matrix elements to perform close a comparison with LDA. From this map of electronic structure, we find that two of the bands undergo a symmetry inversion that is indicative of Mott-Hubbard interactions, while other reflection matrix elements adhere to the single particle first principles LDA wavefunction symmetries. The detailed kinetics of these bands are examined to discuss how they likely contribute to the superconducting wavefunction and energetics of Cooper pairing.

\begin{acknowledgments}

We gratefully acknowledge productive discussions with F. Wang, Y. Ran, A. Vishwanath, T. Tohyama, Z. Tesanovic, I. Mazin and A. Bernevig. R.T. is supported by an SITP fellowship by Stanford University. Use of the Advanced Light Source and Advanced Photon Source was supported by the U.S. Department of Energy (DOE) Office of Science, Office of Basic Energy Sciences (contract no. DE-AC02-05CH11231 and DE-AC02-06CH11357).

\end{acknowledgments}


\begin{thebibliography}
\bibitem{SMo} A linear increase in quasiparticle intensity as binding energy approaches the Fermi level is assumed, as this is a typical qualitative trend for low energy electrons.
\end{thebibliography}
\end{document}